# Contact conductance governs metallicity in conducting metal oxide nanocrystal films

Corey M. Staller[1‡], Stephen L. Gibbs[1‡], Xing Yee Gan[1], Jay T. Bender[1], Karalee Jarvis[2], Gary K. Ong[1], Delia J. Milliron[1]*

[1]McKetta Department of Chemical Engineering, University of Texas at Austin, Austin, Texas 78712, USA.

[2]Texas Materials Institute, University of Texas at Austin, Austin, Texas 78712, USA

‡These authors contributed equally.

*Corresponding author. Email: milliron@che.utexas.edu

**ABSTRACT:** In bulk semiconductor materials, the insulator-metal transition (IMT) is governed by the concentration of conduction electrons. Meanwhile, even when fabricated from metallic building blocks, nanocrystal films are often insulating with inter-nanocrystal contacts acting as electron transport bottlenecks. Using a library of transparent conducting tin-doped indium oxide nanocrystal films with varied electron concentration, size, and contact area, we test candidate criteria for the IMT and establish a phase diagram for electron transport behavior. From variable temperature conductivity measurements, we learn that both the IMT and a subsequent crossover to conventional metallic behavior near room temperature are governed by the conductance of the inter-nanocrystal contacts. To cross the IMT, inter-nanocrystal coupling must be sufficient to overcome the charging energy of a nanocrystal, while conventional metallic behavior requires contact conductance to reach the conductance of a nanocrystal. This understanding can enable the design and fabrication of metallic conducting materials from nanocrystal building blocks.

**MAIN TEXT:** Colloidal nanocrystals are attractive as solution-processible precursors for electronic and optoelectronic devices,(*1–5*) but the conditions for achieving metallic conductivity remain the subject of intense investigation.(*6–9*) In bulk materials, the Mott criterion describes a critical electron concentration at which materials transition from insulating to metallic.(*10*) Metals (above this threshold) have a finite conductivity in the zero-temperature limit, while the resistivities of insulators diverge at low temperature and their conductivity is thermally activated, increasing exponentially with temperature. Conductivity in nanocrystalline films, however, typically remains thermally activated even as the electron concentration exceeds the Mott criterion for an IMT.(*6–9, 11, 12*) For instance, transparent conducting tin-doped indium oxide (ITO) nanocrystals are metallic, owing to a high concentration of dopants, and exhibit strong localized surface plasmon resonance (LSPR) absorption of infrared light.(*13–15*) Nonetheless, even when their insulating ligands are removed, allowing direct contact between nanocrystals, deposited films usually behave as insulators, electrons move by thermally activated hopping, and film resistance diverges at low temperatures.(*1, 11, 12, 16–18*) Conductivity can be increased, ultimately leading to metallicity, by strengthening electronic coupling, e.g., using solution- or vapor-phase deposition to fill the spaces between metal oxide nanocrystals,(*2, 7, 9*) but the criterion for the IMT and the conditions for conventional metallic behavior, where conductivity decreases as temperature increases, remain to be clearly established.

One IMT criterion that has been proposed for nanocrystal films in essence requires contacts that are larger than the characteristic length scale of the conduction electrons to produce a metal, as shown in Eqn. 1, where $k_F$ is the Fermi wavenumber, which increases with electron concentration ($n$) as $n^{1/3}$, and $r_c$ is the contact radius.(*6*)

$$k_F r_c \geq 2 \qquad (1)$$

Experimental tests of this prediction in films of doped silicon or zinc oxide nanocrystals have found qualitative agreement,[6–9] but these tests have been limited by the challenges of tuning $n$ over a wide range and of controllably varying nanocrystal size ($r_{NC}$) and $r_c$.

A second criterion, developed for granular metals and not previously applied to nanocrystal films, predicts an IMT when the conductance of the nanocrystal contacts reaches a critical value ($g^{IMT}$) given by:

$$g^{IMT} = \frac{1}{6\pi} \ln\left(\frac{E_c}{\delta}\right) \propto \ln\left(r_{NC}^2 n^{1/3}\right) \qquad (2)$$

where $E_c$ is the nanocrystal charging energy and $\delta$ is the mean energy spacing at the Fermi level.[19–21] Contact conductance ($g_c$) may be related to $r_c$ via the Sharvin equation, which is suitable for contacts small compared to the mean free path:[22]

$$g_c = \frac{k_F^2 r_c^2}{4} \qquad (3)$$

To meet this criterion, the coupling between nanocrystals (observed as $g_c$) must be sufficient to overcome the transport barriers due to $E_c$. More discussion of Eqn. 2 is found in the Supplementary Materials, Eqns. S11-S17.

Furthermore, a criterion describing necessary conditions for nanocrystal films to exhibit conventional metallic behavior has remained unspecified. Indeed, metals close to the IMT threshold, whether bulk or nanocrystalline materials, often have unusual, *negative* thermal coefficient of resistivity (TCR), exhibiting a logarithmic or power law temperature dependence of conductivity even up to room temperature.[7, 23–25] Fabricating nanocrystal films with conventional metallic behavior (positive TCR) has been difficult[2, 7] and the materials requirements to achieve this behavior have not been established.

Here, by observing the temperature-dependent conductivity of a library of 54 ITO nanocrystal films with varied $n$, $r_{NC}$, and $r_c$, we show that the contact conductance criterion (Eqn. 2) is more predictive of the IMT than Eqn. 1. Contact conductance also plays a central role in determining the crossover to conventional metallic behavior near room temperature, which we find occurs when $g_c$ reaches the nanocrystal conductance, $g_{NC}$. These two criteria together establish a phase diagram that quantitatively describes electron transport and can broadly enable the design and fabrication of metallic conducting materials from nanocrystal building blocks.

**Preparation and characterization of nanocrystal films**

In solution deposited films of ITO nanocrystals, we removed insulating ligands to leave the nanocrystals in direct contact, then tuned $r_c$ by conformal deposition of indium oxide to control the neck between adjacent nanocrystals. Nanocrystals are synthesized with sizes from 5 to 20 nm in diameter with low dispersity by solution-phase synthesis during which Sn dopants are incorporated uniformly throughout the nanocrystals up to 5 at% (Fig. 1A and S1).(*26*) Organic ligands that facilitate solvent casting of uniform thin films are then chemically stripped, leaving the bare nanocrystals in direct contact.(*11*, *27*) Following a strategy used previously for ZnO nanocrystal films,(*7*, *9*) the spaces between the nanocrystals are in-filled by ALD, first with indium oxide to conformally coat the interior surfaces and variably increase $r_c$, then with aluminum oxide to eliminate surface depletion layers that would diminish $g_c$ (Fig. 1B-E).(*11*, *12*) Fabrication of the films, which are highly transparent, is further described in the Supplementary Materials, with a photo and additional electron micrographs in Figs. S2-6. Larger nanocrystals, higher dopant concentration, and increasing indium oxide ALD thickness all tend toward metallicity, but none of these factors is independently predictive of the transport mechanism.

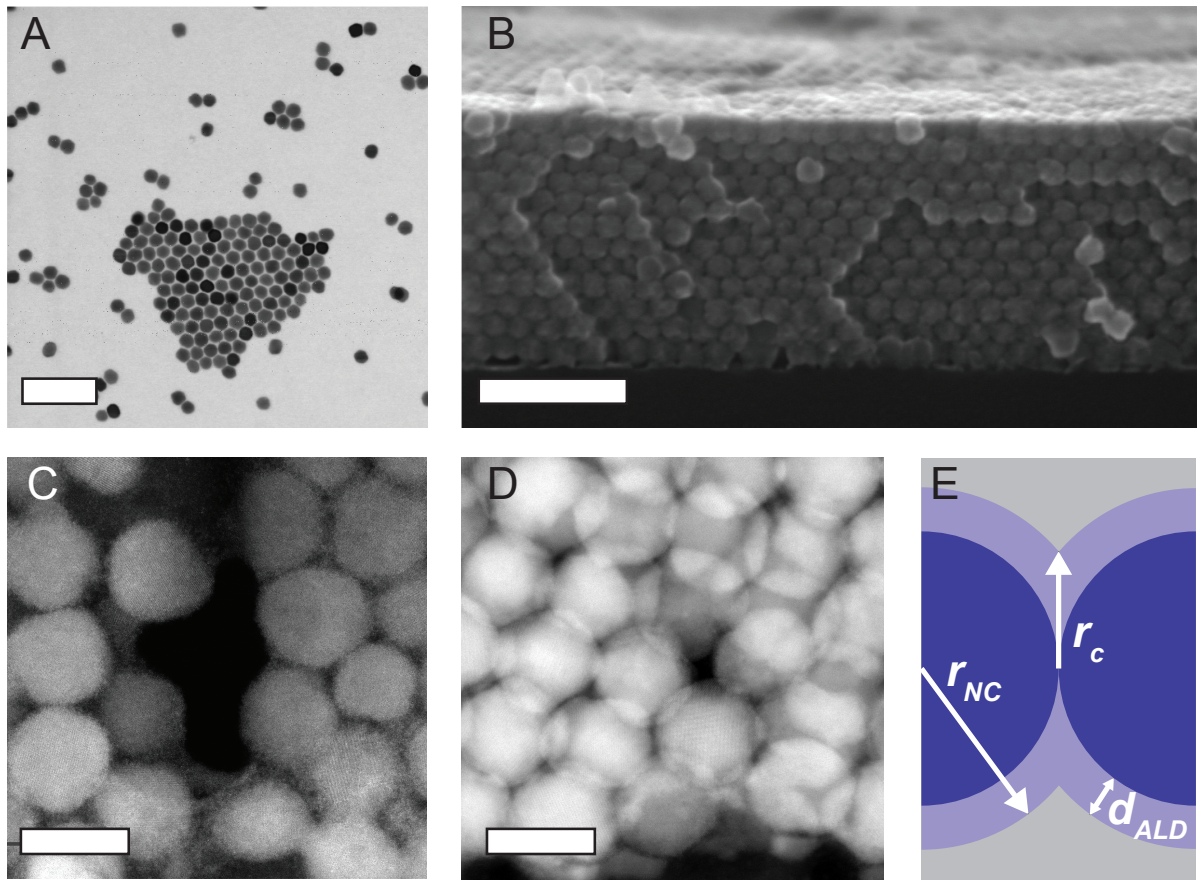

**Fig. 1. Electron microscopy of nanocrystal films.** (**A**) Bright-field scanning transmission electron microscopy (STEM) of 20 nm 3at% Sn ITO nanocrystals. (**B**) Cross-sectional scanning electron microscopy of a film of 20 nm $In_2O_3$ nanocrystals after indium oxide ALD. Scale bars 100 nm. (**C, D**) High-resolution cross-sectional STEM high-angle annular dark field images of finished films with no indium oxide ALD coating (**C**) and 40 cycles of indium oxide ALD (**D**), with scale bars 5 nm. (**E**) Indium oxide ALD (light blue) expands both the nanocrystals (dark blue) and their contact radius with aluminum oxide (gray) filling the remaining pore volume.

## Evaluation of transport mechanisms

The transport mechanism in each of 54 films, made from 9 different batches of nanocrystals with varied cycles of ALD indium oxide, was ascertained by examining the temperature dependence of the conductivity from above room temperature down to a few degrees Kelvin (Fig. 2 and S7). Samples were categorized as insulators or metals, with either conventional (positive TCR) or unconventional (negative TCR) behavior near room temperature, on the basis of the functional dependence of their conductivity on temperature (Fig. 2). Conductivity of insulating samples was thermally activated and could be fit to the Efros-Shklovskii variable-range hopping (ES-VRH) model with a Gaussian dispersion of energy levels (Fig. 2A), similar to many nanocrystal films reported previously.(*6–9, 11, 16, 17, 28*) The -0.8 exponent in the temperature dependence exhibited here is a result of temperature dependent heat capacity of ITO, as described previously.(*11*) Adding indium oxide by ALD increases $r_c$ and the conductivity becomes less strongly temperature dependent, with an apparent finite resistivity in the zero-temperature limit, signaling the IMT threshold has been crossed. These data can be fit to a granular metal or Fermi liquid model, or both in different temperature ranges (Fig. 2B), but they do not exhibit a positive TCR at any temperature.(*21, 21, 29, 30*) This unconventional metallic transport behavior (i.e., negative TCR) is similar to other ALD in-filled ZnO and ITO nanocrystal films.(*7–9, 11*) Finally, a crossover to conventional metallic behavior occurs with a positive TCR around and above room temperature, while diminishing conductivity with decreasing temperature could still be observed at lower temperature (Fig. 2C). A single nanocrystal composition (e.g., 5 at% Sn, 15 nm diameter) can produce films in all three electron transport regimes, depending on the extent of indium oxide ALD (Fig. 2A-C, yellow) while films of large, highly doped nanocrystals (e.g., 5 at% Sn, 20 nm) were unconventional metals as deposited (Fig. 2B, green), their point contacts being sufficient to introduce significant coupling between nanocrystals even without added indium oxide. The models are fully described and fit parameters reported in the Supplementary Materials (Fig. S8 and Tables S1-4).

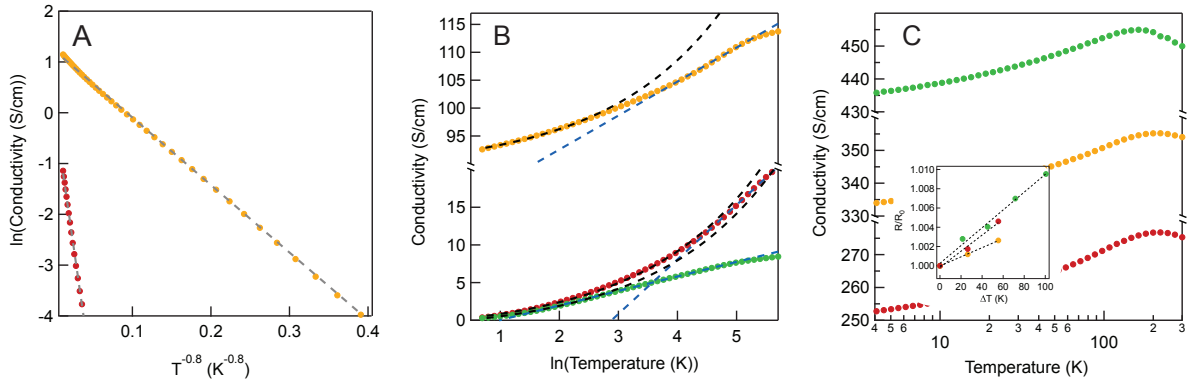

**Fig. 2. Electron transport mechanism determined by variable temperature conductivity.**
(**A**) Example variable temperature conductivity fits (dashed lines) to the ES-VRH model for insulators. (**B**) Unconventional metallic conductivity fits the Fermi liquid (black dashed lines) and granular metal (blue dashed lines) conduction models. (**C**) Conventional metallic behavior is observed around room temperature in some samples. Data are for undoped, 5 nm nanocrystals (red) with 0, 4, or 20 cycles of indium oxide ALD, for 5 at% Sn, 15 nm nanocrystals (yellow) with 0, 8, or 30 cycles, and for 5 at% Sn, 20 nm nanocrystals (green) with 0 or 40 cycles.

## Experimental assessment of IMT criteria

To evaluate which criterion (Eqn. 1 or 2) is most successful in predicting the IMT and discover the conditions leading to conventional metallic behavior in ITO nanocrystal films, the physical parameters $n$, $r_c$, and $r_{NC}$ were determined for each sample. The electron mean free path and initial electron concentration, $n_0$, before ALD coating were determined by fitting LSPR spectra of dispersed nanocrystals (Fig. S9 and Table S5).(*14*) This approach avoids the complication of spectral shifts and broadening induced by LSPR-LSPR coupling in nanocrystal films.(*2*, *13*, *18*, *27*) The effective contact radius, $r_c$, was calculated from $g_c$ values found by analysis of film conductivity. A tunneling or b-contact model was used to find $r_c$ for samples with no indium oxide ALD coating,(*6*, *9*, *11*) while for samples with ALD-enhanced contact area we describe the contact radius using Eqn. 3.(*6*, *22*)

We model the film as a random resistor network,(*28*) with each resistor comprising the resistance of a nanocrystal and the contact resistance $g_c^{-1}$, in series.(*9*) The electronic properties (*n* and the electron mean free path) extracted by fitting LSPR spectra(*14*) are used to calculate the conductance of a nanocrystal, $g_{NC}$, and so determine $g_c$ and $r_c$ for each sample. In this analysis, the thickness of indium oxide added by ALD, $d_{ALD}$, is considered as a single parameter that constrains the relationship between $r_c$ and $r_{NC}$ geometrically (Fig. 1E). The electron concentration used for analysis was reduced from $n_0$ in proportion to the increasing nanocrystal volume and the electron mean free path was adjusted to account for surface scattering, depending on $r_c$ and $r_{NC}$ for each sample.[15] Electron concentrations determined by Hall effect measurements were found to be in good approximate agreement with the optically derived values for metallic films, where Hall analysis is expected to be most reliable (Fig. S10). Further information on establishing these physical parameters is found in the Supplementary Materials.

To evaluate the success of Eqn. 1 in predicting the IMT, we considered all samples and found some exhibiting metallic transport, even one with conventional metallic behavior near room temperature, at values of $k_F r_c$ significantly below the expected threshold value of 2, down to about half this value (Fig. 3A). Based on the cross-validation of electron concentrations by optical and Hall effect analysis, and on consideration of potential errors in the values of $r_c$, we conclude with high certainty that our data violate the predictions of Eqn. 1. In fact, previous literature examining the IMT in ZnO nanocrystal films suggested the precise threshold value of $k_F r_c$ may even be higher than 2;(*8*) if so, the disagreement with our results would be exacerbated.

Alternatively, the IMT can be predicted on the basis of the conductance of the contacts (Eqn. 2), instead of contact size. We find quantitative agreement with all 54 samples satisfying this criterion, including metals found to have $g_c$ as little as 36% above the critical value and insulators found to have $g_c$ only 25% below it (Fig. 3B). Eqn. 1 can also more fundamentally be stated in terms of contact conductance, namely that the IMT occurs when the contact conductance exceeds the quantum conductance ($e^2/\pi\hbar$). But the difference is nontrivial with $g^{IMT}$ in Eqn. 2 depending explicitly on nanocrystal size and the dielectric constants of the nanocrystals and surroundings, and generally having a value well below (0.15 to 0.35 times for our samples, Fig. 3B) the quantum conductance.

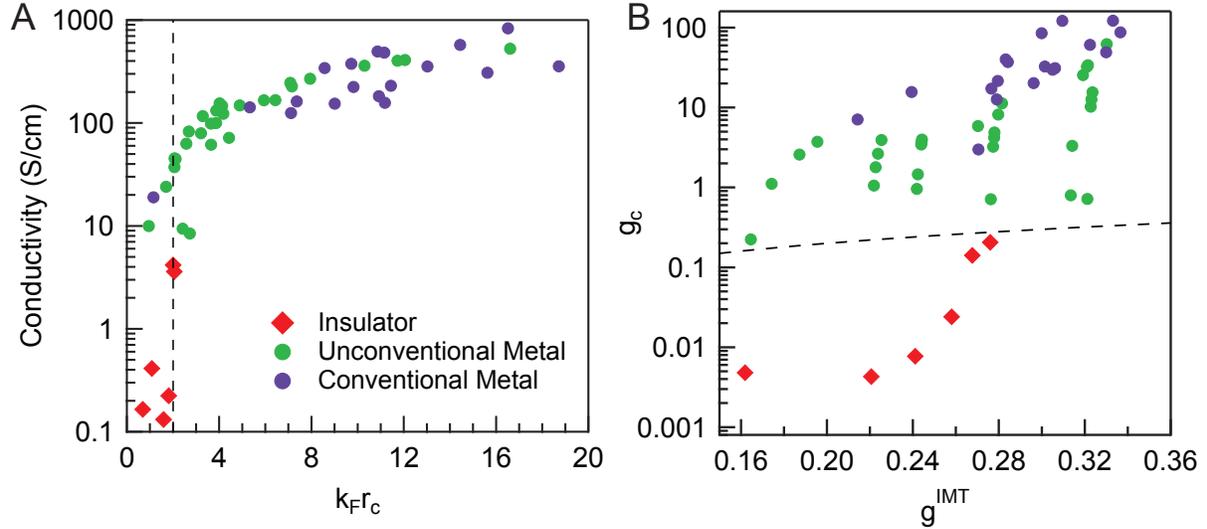

**Fig. 3. Evaluation of candidate IMT criteria.** (**A**) Room temperature conductivity of ITO nanocrystal films with the observed transport mechanism indicated by the symbol for each data point. The vertical dashed line indicates the IMT criterion given by Eqn. 1. (**B**) Comparison to the criterion given by Eqn. 2 shows all samples in quantitative agreement.

**Development of transport mechanism phase diagram**

Samples whose contacts exceed the conductance threshold $g^{IMT}$ may or may not exhibit conventional metallic behavior (positive TCR) and the magnitude of their conductivity fails to predict the transport behavior. For instance, we report metals with room temperature conductivity as high as 500 S cm⁻¹ to have unconventional temperature dependence and metals with room temperature conductivity well below 100 S cm⁻¹ to be conventional metals (Fig. 3A). It was suggested previously(*21*) that metals will cease to behave as granular conductors when $g_c$ is no longer less than $g_{NC}$, which we therefore hypothesized may determine the crossover to conventional metallic conductivity in our ITO nanocrystal films. Indeed, with near perfect agreement, samples with

$$g_c \geq g_{NC} \qquad (4)$$

have a positive TCR near room temperature (Fig. 4A), meaning their conductivity is limited by the conductivity of the nanocrystals themselves, rather than the contacts.

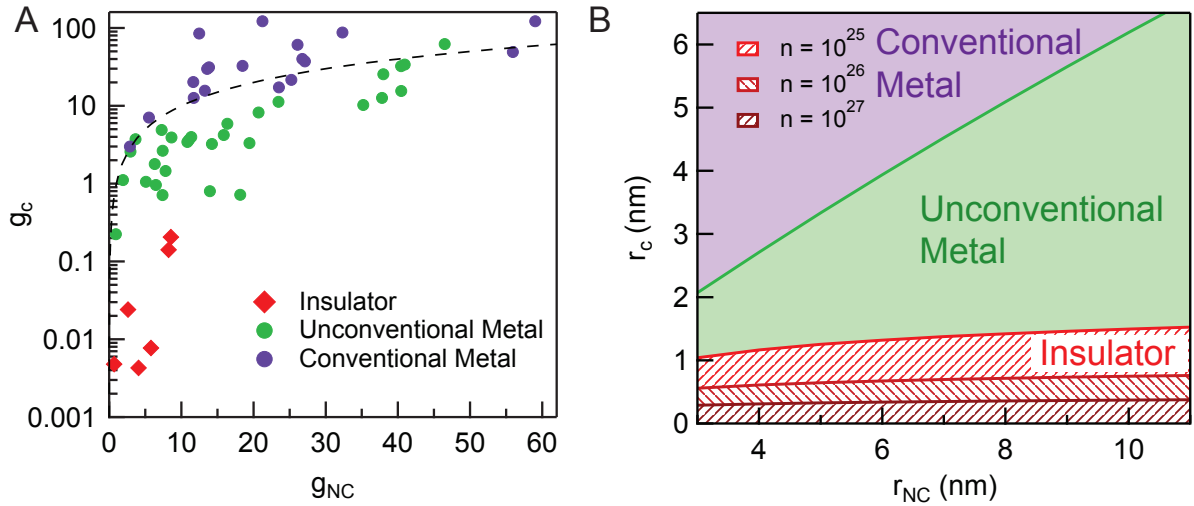

**Fig. 4. Electron transport mechanism phase diagram.** (**A**) Nearly perfect agreement with the criterion $g_c \geq g_{NC}$ to predict conventional metal behavior near room temperature (**B**) The two thresholds (Eqns. 2 and 4) determine the boundaries on the electron transport mechanism phase diagram. Note that the IMT boundary depends on $n$ while the crossover to conventional metal behavior does not.

The two criteria (Eqns. 2 and 4), when expressed in terms of physical characteristics of the nanocrystal films ($n$, $r_c$, and $r_{NC}$), establish boundaries on an electron transport phase diagram (Fig. 4B). The crossover to conventional metallic behavior is independent of $n$, since it depends only on the ratio of $g_c$ and $g_{NC}$, which scale identically with $n$. In contrast, the IMT depends strongly on $n$. For a given nanocrystal size, the required contact area to cross the IMT is reduced at higher $n$, recalling the qualitative lessons of the analysis by Chen, et al.(*6*)

**Discussion**

Variable temperature conductivity measurements of films of nanocrystals varying in size, electron concentration, and contact radius indicate the derived transition and crossover criteria are robust. We have established the electron transport phase diagram for doped metal oxides, which have the potential to be applied as transparent contacts in optoelectronic devices, while the derived results are also expected to apply to other doped semiconductors, including doped silicon nanocrystals,(6) ZnO nanocrystals,(7–9) or metal chalcogenide nanocrystals doped remotely.(3, 4) The criteria also make plain the importance of the contact conductance in governing transport, suggesting novel strategies for design and fabrication of metallic nanocrystal films. For example, highly conductive contacts could be created by introducing distinct, conductive materials at the contact points(31) or by using faceted nanocrystals that pack during deposition to enhance contact area.(2) More broadly, these guidelines could enable the deliberate tuning of nanocrystal films to meet the design criteria for specific electronic applications.

**Author Information**

Corresponding Author

E-mail: milliron@che.utexas.edu

Telephone: (512)232-5702

Notes:

The authors declare no competing financial interest.

**Associated Content**

Supporting information: The Supporting Information is available free of charge on the ACS Publications website at DOI:XXX. Supplementary text describing materials and methods and explaining, in detail, the formulation required to fit data and calculate resulting parameters.

Figures S1-S11 and Tables S1-S5 providing data supporting the analysis and discussion in the main text, including additional electron microscopy, optical spectroscopy, temperature-dependent conductivity, Hall effect, and ellipsometric porosimetry data.

**Acknowledgements:** We thank Drs. X. Li and T. M. Truskett for critical feedback on the manuscript, V. Lakhanpal and Dr. I. R. Gerba-Dolocan for assistance with sample preparation for TEM, and K. Bustillo for advice on TEM imaging conditions. Funding was provided by the National Science Foundation (NSF) through the Center for Dynamics and Control of Materials: an NSF MRSEC grant DMR-1720595 (XYG, DJM), NSF grant CHE-1905263 (SLG, DJM), NASCENT, an NSF ERC (EEC-1160494, CMS), NSF Graduate Research Fellowship DGE-1610403 and DGE-2137420 (JTB), and a Welch Foundation grant F-1848 (GKO, DJM).

**Author contributions**: Conceptualization, sample fabrication, and data acquisition was carried out by CMS, GKO, XYG, JTB, and KJ. Analysis and visualization were performed by CMS and SLG. Writing for the original draft was completed by CMS, SLG, and DJM. Review and editing were carried out by CMS, SLG, XYG, JTB, KJ, GKO, and DJM.

**For TOC only:**

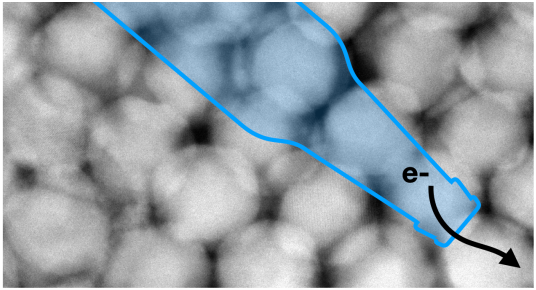

# Supporting Information

# Contact conductance governs metallicity in conducting metal oxide nanocrystal films

**Authors:** Corey M. Staller[1‡], Stephen L. Gibbs[1‡], Xing Yee Gan[1], Jay T. Bender[1], Karalee Jarvis,[2] Gary K. Ong[1], Delia J. Milliron[1]*

**Affiliations:** [1]McKetta Department of Chemical Engineering, University of Texas at Austin; Austin, Texas 78712, USA. [2]Texas Materials Institute, University of Texas at Austin, Austin, Texas 78712, USA

‡These authors contributed equally.

*Corresponding author. Email: milliron@che.utexas.edu

**This PDF file includes:**

Supplementary text describing materials and methods and explaining, in detail, the formulation required to fit data and calculate resulting parameters

**Supplementary Text**

<u>Materials and Methods</u>

**Slow injection colloidal synthesis of tin-doped indium oxide (ITO) nanocrystals.** A detailed explanation of synthesis can be found in prior work (Ref. *15* in the main text). In brief, indium acetate was dissolved in oleic acid at $100°C$ and degassed for an hour. Then the solution was brought to $150°C$ and left to react under nitrogen for at least two hours to form indium oleate precursor. This precursor solution is then added dropwise into a 100 mL three-neck flask containing a hot bath of oleyl alcohol ($290°C$). After the injection is complete, the nanocrystals are purified by dispersing in hexane, flocculating with ethanol, centrifuging, and redispersing in hexane. The size of the nanocrystals is controlled by the injection volume. The tin doping concentration is controlled by substituting the desired fraction of tin acetate with that of indium acetate when forming the metal oleate precursor.

**Inductively coupled plasma atomic emission spectroscopy (ICP-AES).** For each sample, an aliquot of known volume was dried and digested in aqua regia. The dissolved metal ions were then diluted to 2% v/v acid and loaded into the Varian 720-ES ICP-AES. The intensity of the optical emission of tin and indium were compared to known values from prepared standards (Sigma Aldrich TraceCERT® 1000 mg In/L nitric acid and 1000 mg Sn/L nitric acid). This enabled us to calculate the ratio of tin to indium atoms as well as to know the concentration of nanocrystals in each sample (taking the density of indium oxide to be 7.14 g/mL). It is necessary to know the concentration of nanocrystals to do quantitative fitting of the LSPR peak.(*3*)

**Preparation of bare nanocrystal films.** Nanocrystal dispersions in 1:1 hexane:octane by volume and of concentration ~60 mg/mL solvent were prepared as the spin solution. 30 μL of the spin solution was dropped onto a quartz substrate and immediately spun at 1000 rpm for 30 seconds and then 4000 rpm for 60 seconds. This results in visibly transparent films (Fig. S5) of ~200 nm in thickness. To remove the native organic ligands, the films are let to soak in a solution of 0.05 M formic acid in acetonitrile for 30 minutes to exchanged oleate for formate.(*4*) Then the films are rinsed with neat acetonitrile and annealed under nitrogen at 300°C for 1 hour to vaporize the formate and leave the nanocrystal surface free of organic ligand.

**Atomic layer deposition (ALD).** A detailed explanation of the ALD procedures can be found in prior work.(*4*) In brief, indium oxide was deposited onto bare nanocrystal films in a Cambridge Nanotech Savannah S100 by alternating cycles of 0.1 s pulse of cyclopentadienyl indium, 30 s purge, 0.1 s pulse of water, and 30 s purge. Deposition cycles were determined by the desired deposition thickness. Samples were unloaded for ellipsometric porosimetry. After characterization, all films were coated with 40 cycles of trimethylaluminum (TMA) and water to completely infill with alumina. In the same prior work,(*4*) it was found using time-of-flight secondary ion mass spectrometry (TOF-SIMS) that this procedure resulted in ALD coatings that penetrate the full thickness of the film.

**Scanning transmission electron microscopy (STEM) and scanning electron microscopy (SEM).** Stock dispersions for samples in this work were diluted to ~0.1 mg/mL in hexane and dropped onto a TEM grid (Ultrathin Carbon Type A, 400 Mesh) for bright field STEM.

Micrographs were captured on a Hitachi S-5500 STEM with 30 kV accelerating voltage (Fig. S1). Cross-sectional micrographs were also captured in the Hitachi S-5500 STEM. To view the topology and thickness of our nanocrystal films, the images were collected in secondary electron (SE) mode (Fig. S6).

**Lamella preparation for transmission electron microscopy (TEM) imaging of cross-sections of nanocrystal films.** The TEM lift-out lamellae were prepared using a Scios 2 HiVac dual-beam FIB/SEM system equipped with an EasyLift nanomanipulator and gas injection systems for carbon and platinum depositions. The nanocrystal films were placed on SEM pin stubs with conductive carbon tape. The TEM cross-sectional lamellae were prepared using the following procedure. First, a 200 nm C protective layer followed by a 200 nm Pt protective layer were deposited on top of the films by E-beam at 5 kV and 1.6 nA. Then, an extra 2.2 μm thick Pt layer was deposited in I-beam at 30 kV and 300 pA. The lamella was then milled from the bulk film at 30 kV and 5 nA, with an intermediate and undercut at 30 kV and 1 nA. The lamella was attached to the side of a lift-out grid post by Pt welding at 30 kV and 0.1 nA. Next, the lamella was thinned down to approximately 80-100 nm using a 100 pA beam, and finally polished at 5 kV and 77 pA at +/- 3 off tilt. The polishing was done until the protective layer was completely removed, resulting in an approximately 15 nm thin lamella as measured by EELS. The lamella was then transferred to the TEM instrument for examination.

**TEM cross-section imaging.** The prepared nanocrystal cross-sections were imaged with a JEOL NeoARM equipped with a 5[th]-order probe spherical aberration corrector. Dark field STEM images were captured with an accelerating voltage of 80 kV using a JEOL high angle annular dark field detector (HAADF).

**Spectroscopy of colloidal dispersions.** Ligand-capped nanocrystals were dispersed in tetrachloroethylene at a known (from ICP-AES) concentration. This dispersion was loaded into a liquid cell consisting of two KBr plates and a 0.05 mm spacer. Depending on the anticipated position of the plasmon peak, the liquid cell was then loaded into either a Bruker Vertex 70 Fourier transform infrared (FT-IR) spectrophotometer or an Agilent Cary series ultraviolet-visible-near infrared (UV-vis-NIR) spectrophotometer in transmission mode.

**LSPR peak fitting.** To obtain electronic parameters such as the electron density $n$ and bulk mean free path $l_{bulk}$, dispersion phase spectra are fit to the heterogeneous ensemble Drude approximation (HEDA) model (Fig. S9, Table S5).$(3)$ Fit results for samples used in this work that are not shown here can be found in Ref 15 in the main text. We note that the LSPR signal is too weak to fit for samples 0 at% Sn 5 nm and 0 at% Sn 15 nm. In that case, we determined $n$ and $l_{bulk}$ values based on similar samples fit in Ref 15 in the main text.

**Ellipsometric Porosimetry (EP).** To obtain the volume of pores within the nanocrystal films, we performed EP. A detailed explanation of the procedure can be found in prior work.$(4)$ The film is loaded into a J. A. Woollam Environment Cell that controls the partial pressure of solvent vapor, toluene in our case. As the partial pressure increases, the volume fraction of solvent in the nanocrystal film is tracked by fitting the optical response using a J. A. Woollam M-2000 Spectroscopic Ellipsometer. The maximum value of volume fraction reveals the relative pore volume (Fig. S11).

**Conductivity measurements.** For all 54 nanocrystal films, room temperature conductivity was collected on an Ecopia Hall effect measurement system (HMS-5000) using 4-point probe Van der Pauw geometry. Separately, temperature dependent conductivity was collected between 2 K and 300 K (Fig. S7) in a Quantum Design Physical Properties Measurement System with external electronics sources. Current from -1 to 1 uA in 500 nA increments was applied through adjacent corners of the sample using a Keithley 220 current source and the voltage measured on the other two corners using a Keithley 2182 nanovoltmeter. A Keithley 2700 switch box was used to perform 4-point measurements in both configurations and ohmic contact in 4-point and 2-point configurations was ensured at 300 K and the lowest temperature measured.

<u>Calculation of contact radius $r_c$ for 0x indium oxide ALD (b-contact)</u>
For the films where no indium oxide ALD has been deposited, the expected physical contact radius for perfect spheres would be simply a point contact. For this case, $r_c$ is described by a $b$ contact:

$$r_c = \sqrt{r_{NC,0} b} \qquad\qquad \text{Equation S1}$$

Here, $r_{NC,0}$ is the radius of the nanocrystal as deposited, before ALD, and $b$ is the wavefunction decay length, which depends on the nanocrystal radius and the work function of the surrounding medium – in this case alumina.(*1*)

<u>Calculation of $r_c$ for > 0x indium oxide ALD</u>
We set up a system of two equations to solve for two unknowns: $d_{ALD}$, the effective thickness of indium oxide ALD, and $r_c$, the contact radius (See main text Figure 1E for schematic). The first equation relates $r_c$, $d_{ALD}$, and $r_{NC}$ according to geometry of two uniform spheres growing in to one another:(*2*)

$$r_c = \sqrt{d_{ALD}^2 + 2r_{NC}d_{ALD}} \qquad\qquad \text{Equation S2}$$

Then, we developed the second equation by first employing the links and nodes model. Given $r_{NC}$ and initial porosity, $\varphi$, we calculated the bond resistance for a given room temperature conductivity, $\sigma$:

$$R_{bond} = \frac{e^2(\varphi - \varphi_0)^{1.9}}{h\sigma r_{NC}} \qquad\qquad \text{Equation S3}$$

We take the percolation threshold, $\varphi_0$ to be 0.2. The bond resistance can be split into two contributions: one from the conductance of the contact and one from the conductance of the nanocrystal itself:

$$R_{bond} = \frac{1}{g_c} + \frac{1}{g_{NC}} \qquad\qquad \text{Equation S4}$$

Where $g_c$ is defined by the Sharvin equation:

$$g_c = \frac{k_F^2 r_c^2}{4}$$

Equation S5

With the Fermi wavenumber set by the concentration of free electrons, $n$:

$$k_F = (3\pi^2 n)^{1/3}$$

Equation S6

And the intra-nanocrystal conductance, $g_{NC}$ is defined by the nondimensionalized Drude conductivity:

$$g_{NC} = \frac{4r_{NC}nq^2}{3m_e^*\Gamma} * \frac{h}{q^2}$$

Equation S7

Where $m_e^*$ is the electron effective mass ($0.4m_e$ for ITO) and the rate of electron scattering includes contributions from the bulk, the nanocrystal surface, and the cylindrical neck forming between adjacent nanocrystals:

$$\Gamma = \frac{(3\pi^2 n)^{1/3}\hbar}{m_e^*}\left(\frac{1}{\frac{4}{3}r_{NC}} + \frac{1}{l_{bulk}} + \frac{1}{r_c}\right)$$

Equation S8

Where $l_{bulk}$ is the mean free path of an electron. Therefore,

$$g_{NC} = \frac{4hr_{NC}\pi^{1/3}n^{2/3}}{3^{4/3}}\left(\frac{1}{\frac{4}{3}r_{NC}} + \frac{1}{l_{bulk}} + \frac{1}{r_c}\right)^{-1}$$

Equation S9

Plugging Equations S4-9 into Equation S3, we arrive at the second equation in our system of two equations. Then we iteratively solve to converge on a solution for $r_c$ and $d_{ALD}$.

Reduction in $n$ with $In_2O_3$ ALD

Based on the calculations above, the growth in $r_{NC}$ with ALD results in a diluted carrier concentration from that measured in dispersion. The new carrier concentration, $n_{new}$, for the larger nanocrystal is calculated with the following formula:

$$n_{new} = \frac{V_0 n_0}{V_{new}} + \left(1 - \frac{V_0}{V_{new}}\right)n_{IO}$$

Equation S10

Where $V_0$ and $V_{new}$ are the volumes of the original nanocrystal and new nanocrystal, respectively, $n_0$ is the carrier concentration of the original nanocrystal, and $n_{IO}$ is the carrier concentration of the deposited indium oxide, which we take to be $7.1\times10^{19}$ cm$^{-3}$ based on prior work.(3)

Criterion for insulator-metal transition (IMT)

When the coupling between nanocrystals meets or exceeds the charging energy, the conductivity of the nanocrystal film will transition from hopping conduction mechanism (insulator) to a granular metal conduction mechanism (metal). This transition occurs when $g_c$ reaches the critical conductance for the activated metal transition, $g^{IMT}$:

$$g^{IMT} = \frac{\ln\left(\frac{E_C}{\delta}\right)}{6\pi} \qquad \text{Equation S11}$$

Where

$$E_C = \frac{e^2}{2\pi\varepsilon_c\varepsilon_0 a} \qquad \text{Equation S12}$$

And where $e$ is the electron charge, $\varepsilon_0$ is the permittivity of vacuum, $a$ is the metallic grain diameter, and

$$\varepsilon_c = \frac{\pi}{2}\varepsilon_m\left(\frac{2\epsilon}{\pi\varepsilon_m}\right)^{\frac{2}{5}} = 1.99 \qquad \text{Equation S13}$$

Here, $\varepsilon_m$ is the dielectric constant of the surrounding medium and $\epsilon$ is the static dielectric constant of the nanocrystal (9 for ITO). Also, the mean energy spacing in a single grain,

$$\delta = \left(g_{E_F}V_{NC}\right)^{-1} \qquad \text{Equation S14}$$

Where $V_{NC}$ is the NC volume and the density of states at the Fermi level is

$$g_{E_F} = \frac{3^{\frac{1}{3}}m_e^*n^{\frac{1}{3}}}{\hbar^2\pi^{\frac{4}{3}}} \qquad \text{Equation S15}$$

This simplifies to Equation 2 in the main text:

$$g^{IMT} = \ln\left(Cr_{NC}^2 n^{\frac{1}{3}}\right) \qquad \text{Equation S16}$$

$$C = \frac{e^2 m_e^*}{3^{\frac{2}{3}}\pi^{\frac{4}{3}}\varepsilon_c\varepsilon_0\hbar^2} \qquad \text{Equation S17}$$

The relationship between $r_c^{IMT}$, $r_{NC}$, and $n$ (plotted as boundaries on the phase diagram in main text Figure 4B) is represented by:

$$r_c^{IMT} = \frac{A}{n^{\frac{1}{3}}\ln\left(Br_{NC}^2 n^{\frac{1}{3}}\right)} \qquad \text{Equation S18}$$

Where $A$ and $B$ are constant coefficients.

Criterion for crossover to conventional metallic behavior near room temperature

We categorize samples as exhibiting conventional metallic behavior when they exhibit a positive temperature coefficient of resistance (TCR) near room temperature. The crossover from unconventional (negative TCR) to conventional metallic behavior occurs when the contact conductance, $g_c$, meets or exceeds intra-nanocrystal conductance, $g_{NC}$. Notably, these two values are equivalently proportional to the free electron concentration ($\propto n^{\frac{2}{3}}$) and therefore this transition is independent of $n$.

Classification of conduction mechanism by fits to the variable temperature conductivity.

*Insulators: Efros-Shklovski variable-range hopping for a Gaussian dispersion of energy levels (ES-VRH-GD)*

The temperature dependence of hopping conduction as defined by Efros-Shklovski variable-range hopping is:

$$\sigma = \sigma_0 \exp\left(-\left(\frac{T_0}{T}\right)^m\right) \qquad \text{Equation S19}$$

Where the characteristic temperature is defined as:

$$T_0 = \left(\frac{3.15e^4}{16\pi^2\varepsilon^2\varepsilon_0^2 k_B Ca^2}\right)^{\frac{1}{3}} \qquad \text{Equation S20}$$

And the exponent, $m$, was found to be ~0.8 through Zabrodskii analysis.(*4*) Fits are plotted in main text Figure 2A and again in more detail in Fig. S9. Fit parameters are reported in Table S2.

*Metals: Granular metal conduction*

Logarithmic temperature dependence of conductivity in nanocrystal films has been attributed to the granular metal conduction mechanism.(*5–7*)

$$\sigma = \sigma_0\left(1 + \frac{1}{2\pi g_T d}\ln\left(\frac{k_B T}{g_T E_C}\right)\right) \qquad \text{Equation S21}$$

Where

$$\sigma_0 = g_T\left(\frac{2e^2}{\hbar}\right)a^{2-d} \qquad \text{Equation S22}$$

Where $g_T$ is the non-dimensional tunneling conductance and $d$ is the system dimensionality. Slope and intercept values were fit to the following linear equation constructed from Equations S21 and S22.

$$\sigma = A_{GM}\ln(T) + B_{GM} \qquad \text{Equation S23}$$

Where

$$A_{GM} = \frac{\sigma_0}{2\pi g_T d} \qquad\qquad \text{Equation S24}$$

$$B_{GM} = \sigma_0 \left( 1 + \frac{1}{2\pi g_T d} \ln\left( \frac{k_B}{g_T E_C} \right) \right) \qquad\qquad \text{Equation S25}$$

Fits are plotted in main text Figure 2B and again in more detail in Fig. S8B and C. Fit parameters are reported in Table S2.

*Metals: Fermi liquid conduction*
The granular metal model is valid for $T < \Gamma$, where:

$$\Gamma = \frac{g_T \delta}{k_B} \qquad\qquad \text{Equation S26}$$

For $T > \Gamma$, the temperature dependence of conductivity is better described by the Fermi liquid model.(*8*)

$$\sigma = \sigma_0 \left( 1 + \frac{1}{2\pi g_T d} \ln\left( \frac{k_B \delta}{E_C} \right) + \frac{1.83}{12\pi^2 g_T} \sqrt{\frac{T}{\Gamma}} \right) \qquad\qquad \text{Equation S27}$$

Where, again, $\sigma_0$ is defined as Equation S22. Slope and intercept values were fit to the following linear equation constructed from Equations S26 and S27.

$$\sigma = A_{FL} T^{1/2} + B_{FL} \qquad\qquad \text{Equation S28}$$

Where

$$A_{FL} = \frac{1.83 \sigma_0}{12\pi^2 g_T \sqrt{\Gamma}} \qquad\qquad \text{Equation S29}$$

$$B_{FL} = \sigma_0 \left( 1 + \frac{1}{2\pi g_T d} \ln\left( \frac{k_B}{g_T E_C} \right) \right) \qquad\qquad \text{Equation S30}$$

Fits are plotted in main text Figure 2B and again in more detail in Fig. S8B and C. Fit parameters are reported in Table S3.

*Metals: Conventional metallic behavior*
Near room temperature, a significant number of our nanocrystal films exhibit classical metal conductivity – a positive temperature coefficient of resistance $\alpha$, where

$$\rho = \rho_0 \left( 1 + \alpha \Delta T \right) \qquad\qquad \text{Equation S31}$$

Where $\rho$ is the film resistivity at a measured temperature, $\rho_0$ is the film resistivity at a reference temperature, and $\Delta T$ is the difference between the measured and reference temperature. Fits are plotted in main text Figure 2C and again in more detail in Fig. S8C. Fit parameters are reported in Table S4.

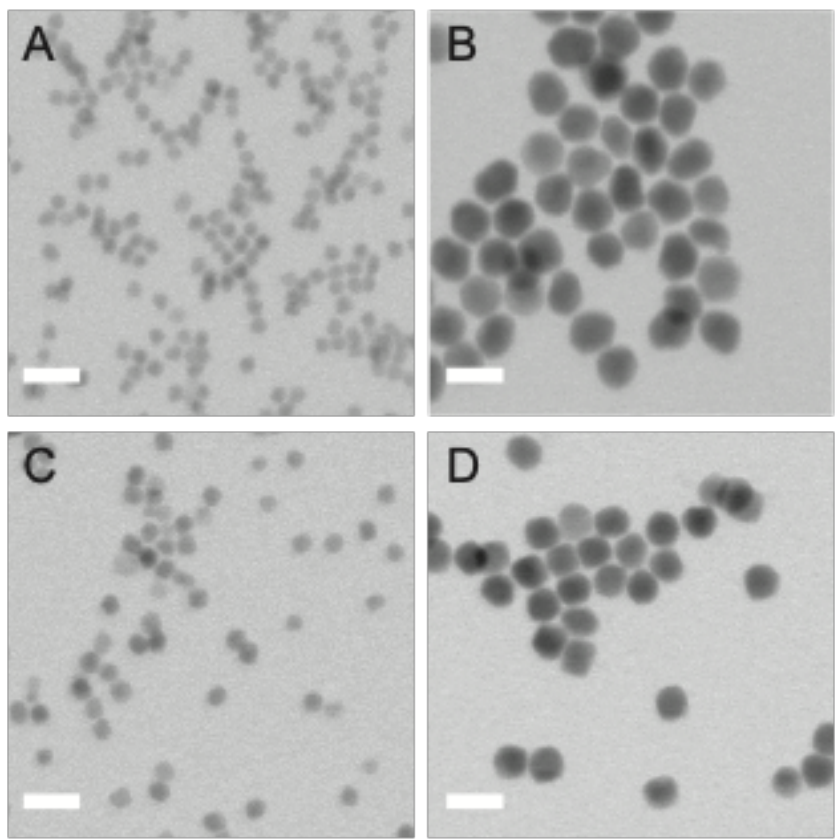

**Fig. S1. STEM micrographs of indium oxide and ITO nanocrystals.**
To obtain size and morphology, bright field (BF) STEM micrographs were collected for samples of the following nominal doping and size: 0 at% 5 nm (A), 0 at% 15 nm (B), 3 at% 5 nm (C), and 3 at% 15 nm (D). Scale bars represent 20 nm. STEM micrographs for the remaining samples used in this work can be found in Ref 15 in the main text.

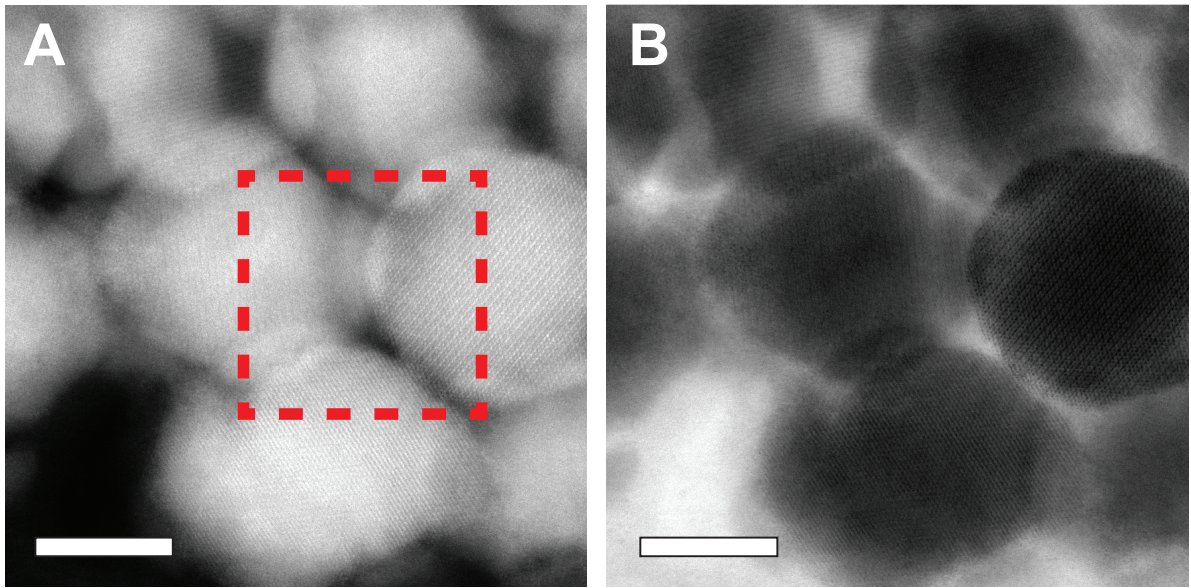

**Fig. S2. High-resolution STEM micrographs of 20 nm 3at% Sn:In₂O₃ nanocrystal film cross-sections.**

Wider area micrographs for the cropped image shown in main text Figure 1D (40 cycles of indium oxide ALD) in HAADF (A) and BF (B), with the dashed red line showing the cropped portion. Scale bars represent 10 nm.

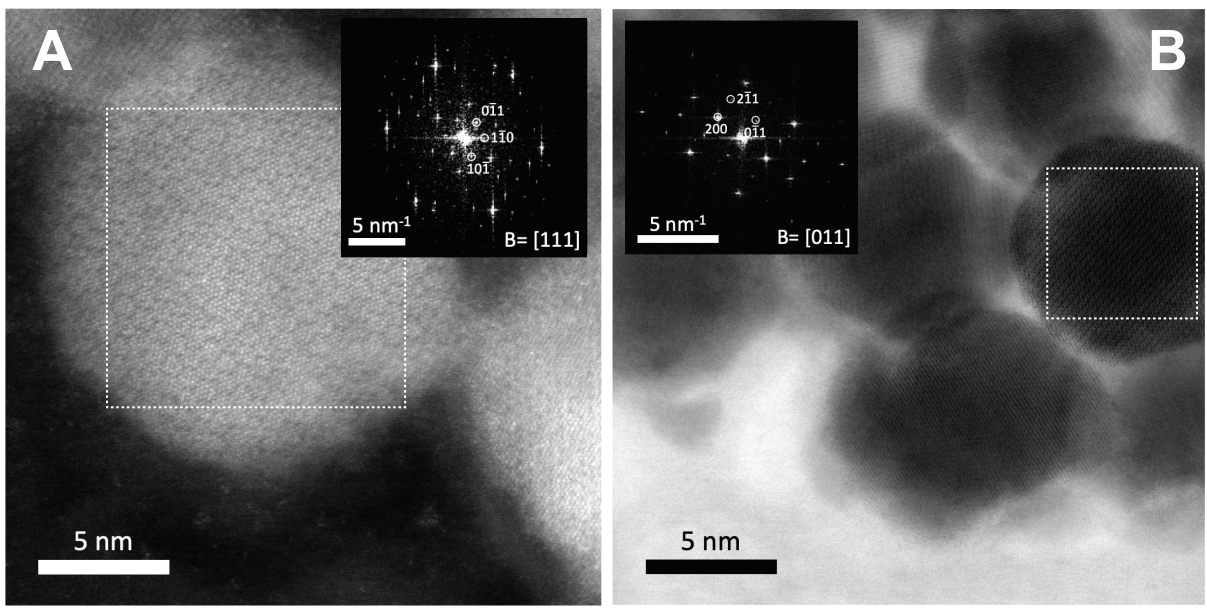

**Fig. S3. STEM micrographs for Fast Fourier Transform (FFT) indexing.**
Using Main Text Fig. 1C (A) and the BF version of Main Text Fig. 1D (B), we indexed an FFT pattern and found the zone axis. The indexed area is outlined in a dashed white box and the FFT pattern is shown in the black insets. The zone axes were found to be [111] and [011] for these selected areas.

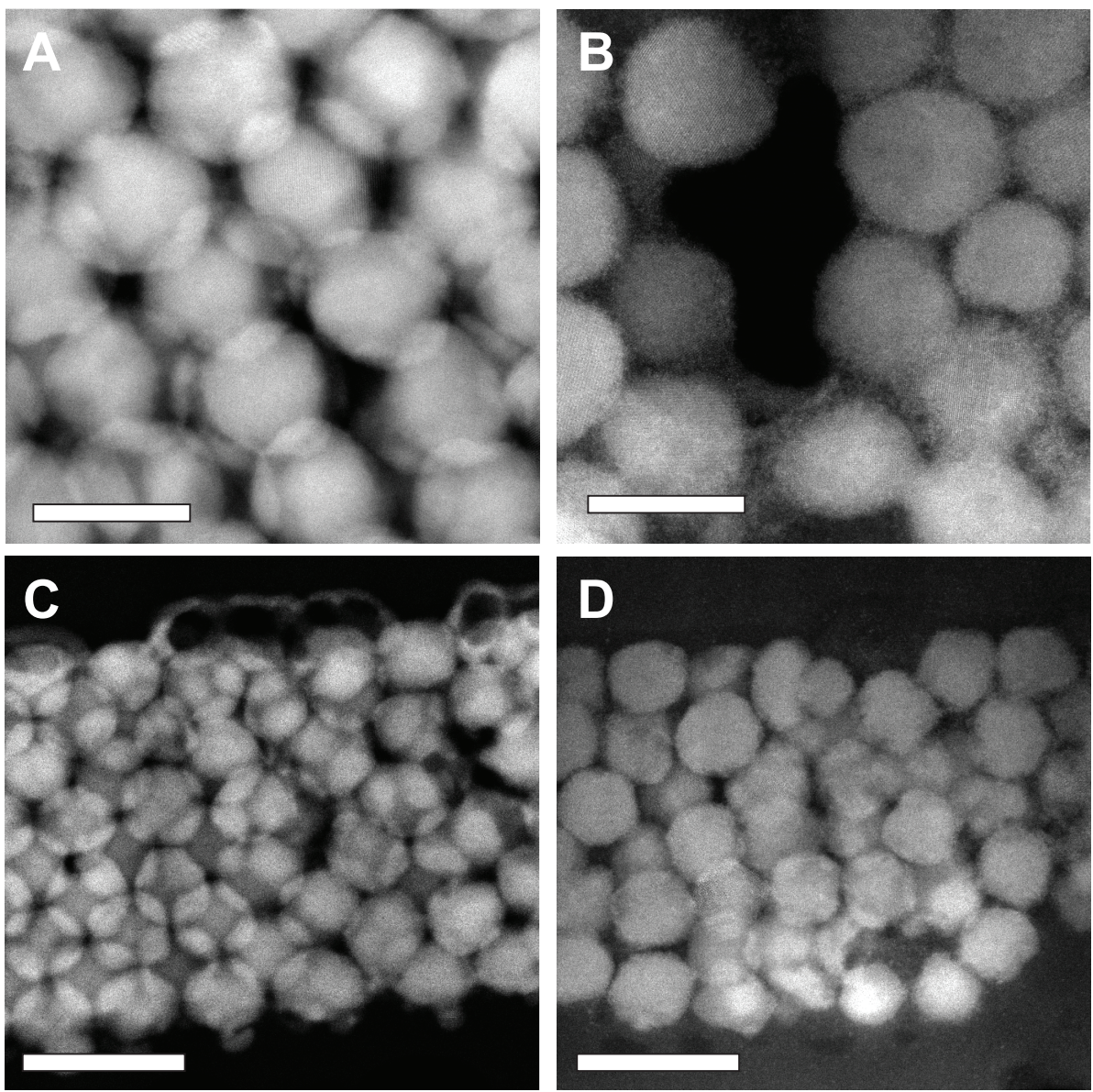

**Fig. S4. More high-resolution STEM micrographs of 20 nm 3at% Sn:In$_2$O$_3$ nanocrystal film cross-sections.**
(A,B) Low magnification HAADF cross-section images for films with 40 cycles (A) and 0 cycles (B) of indium oxide ALD. Scale bars represent 20 nm. (C,D) Lowest magnification HAADF cross-section images for films with 40 cycles (C) and 0 cycles (D) of indium oxide ALD. Scale bars represent 40 nm.

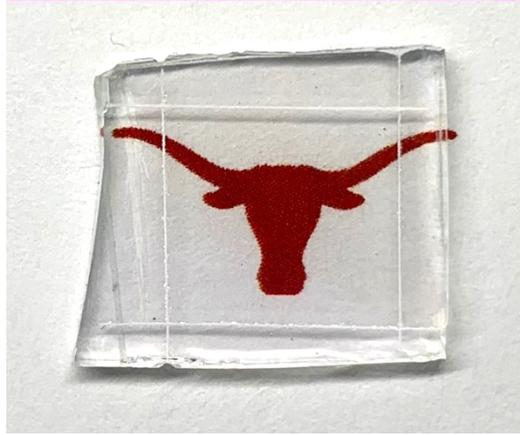

**Fig. S5. Visible transparency of ITO films.**

The image above was taken through a film composed of 3 at% Sn ITO nanocrystals with four cycles of indium oxide ALD.

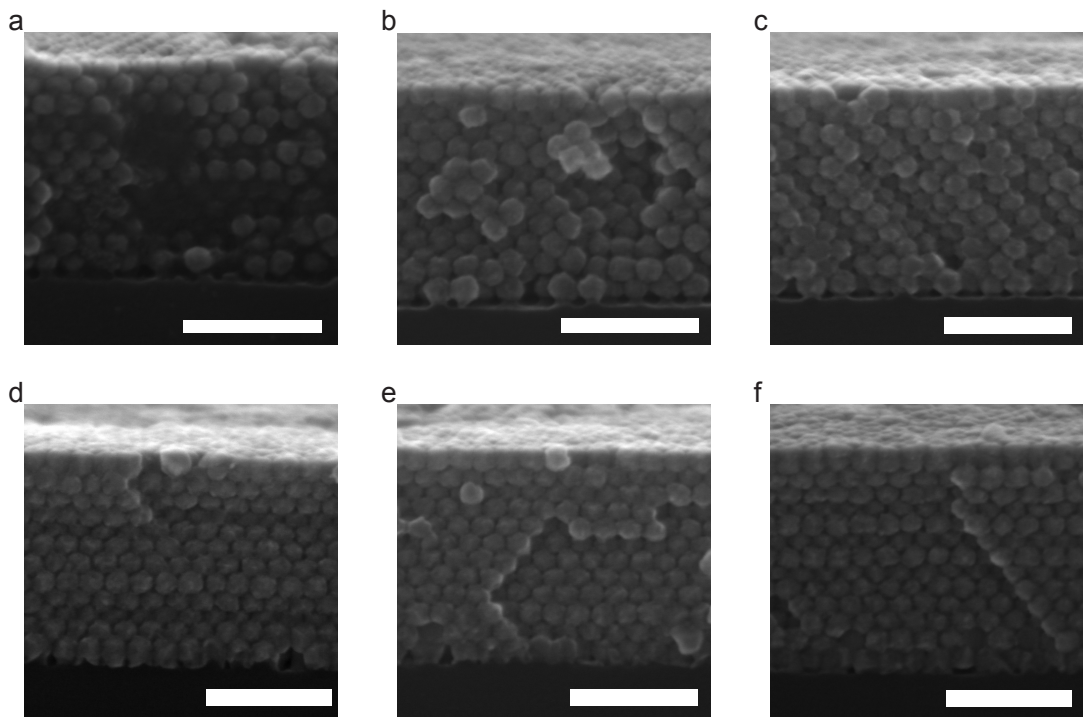

**Fig. S6. SEM cross-sections of nanocrystal films.**
The films are comprised of 20 nm 0 atomic% Sn ITO NCs following 0 (a), 8 (b), 16 (c), 24 (d), 30 (e), and 40 (f) cycles of indium oxide ALD. Scale bars represent 100 nm.

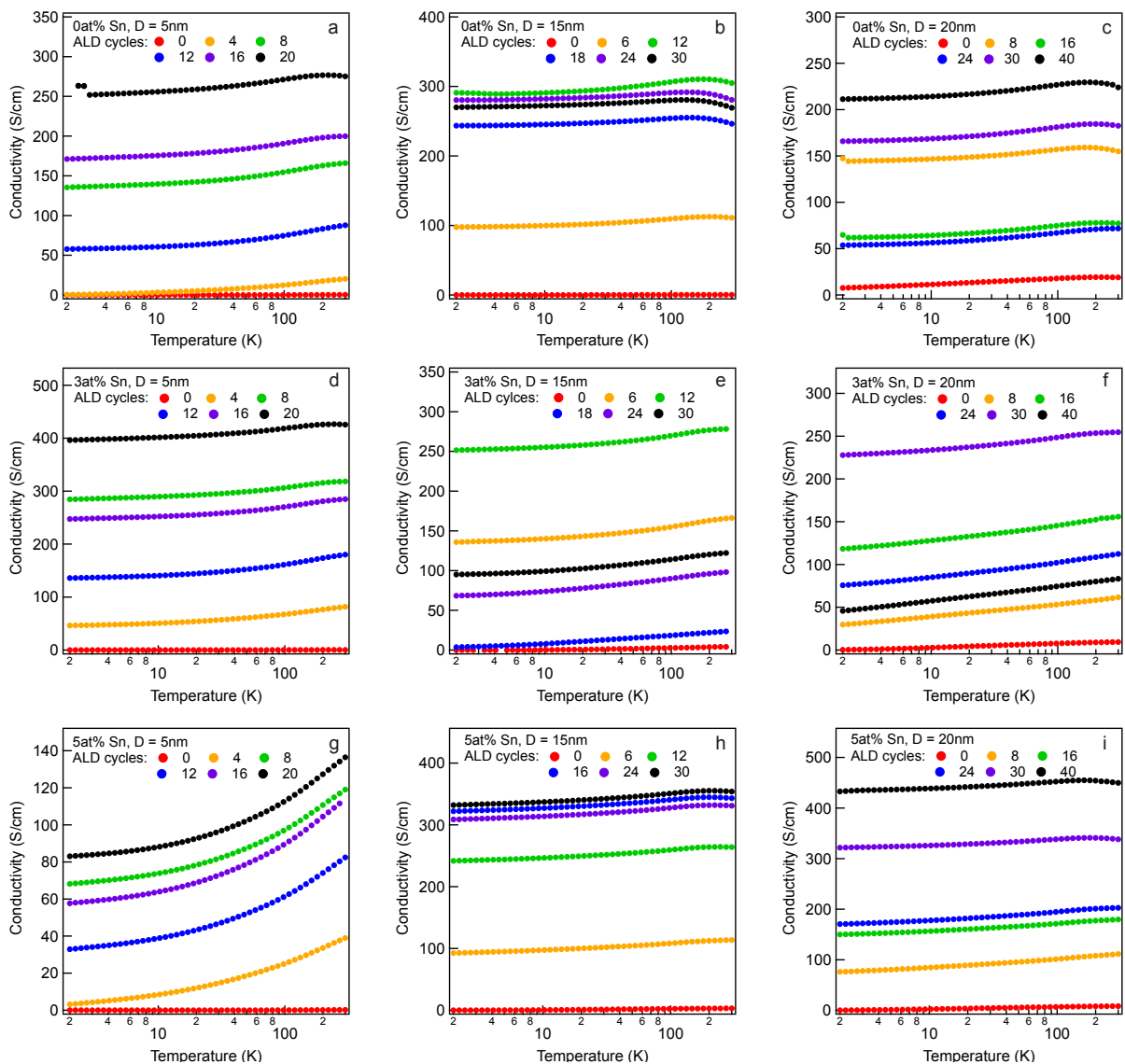

**Fig. S7. Temperature-dependence of conductivity.**
Variable temperature (~2-300K) conductivity for all 54 nanocrystal films that were fabricated for this work.

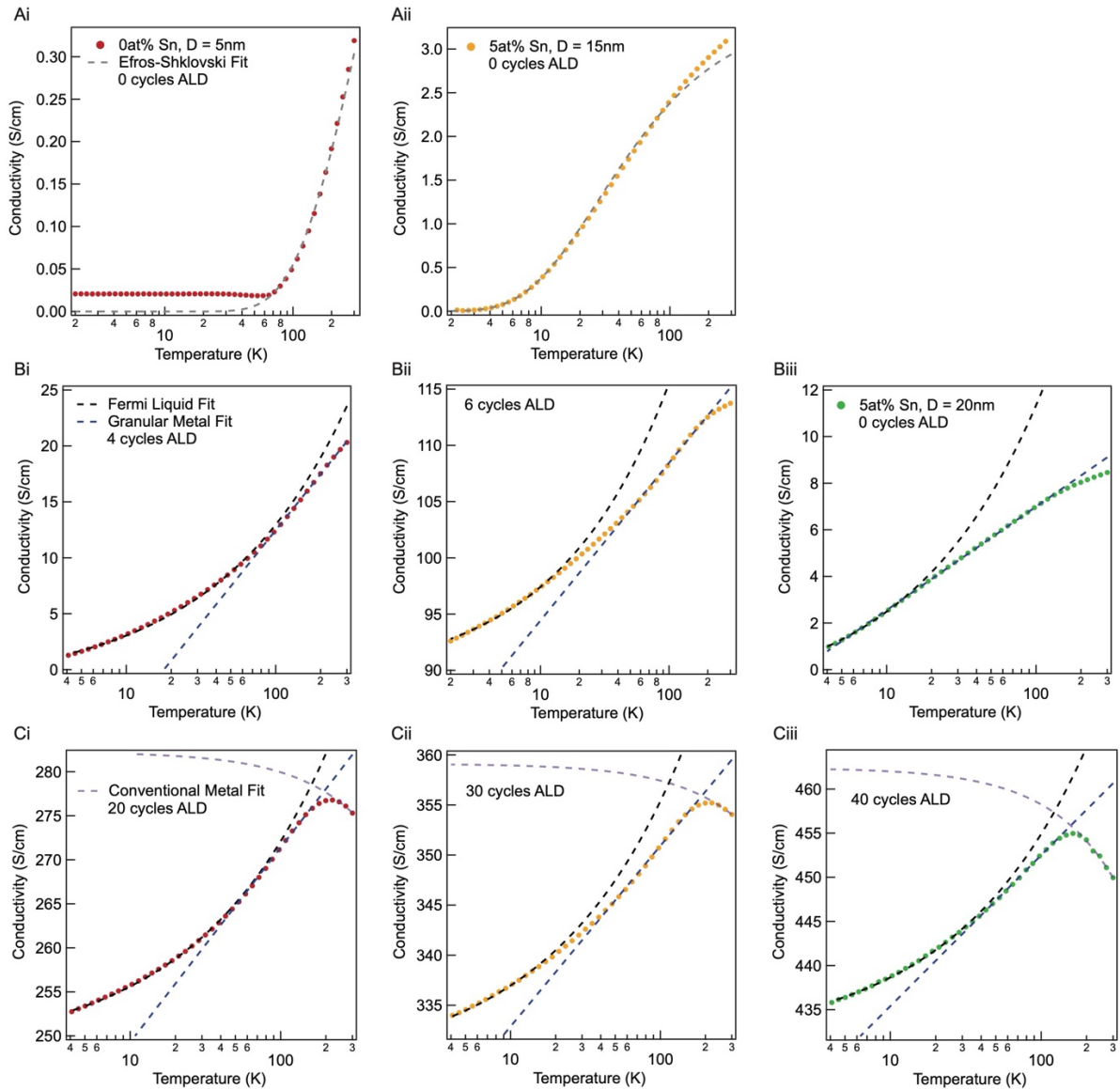

**Fig. S8. Fits to variable temperature conductivity.**
All fits to the variable temperature conductivity data shown in main text Figure 2. Some samples exhibit multiple conduction mechanisms between 2 and 300K; however, no two conduction models were ever fit to data within the same temperature range. The corresponding fit parameters are reported in Tables S1-4.

| Sn doping [%] | nominal diameter [nm] | In$_2$O$_3$ ALD cycles | $a$ [$nm$] | $\sigma_0$ [S/cm] |
|---|---|---|---|---|
| 0 | 5 | 4 | 0.28 | 1.1 |
| 5 | 15 | 6 | 14.3 | 3.4 |

**Table S1. Fit parameters acquired from fitting temperature dependence of conductivity for insulators fit to a ES-VRH-GD conduction mechanism.**

$a$ is the localization length and $\sigma_0$ is the pre-factor in Equation S19.

| Sn doping [%] | nominal diameter [nm] | In$_2$O$_3$ ALD cycles | $A_{GM} \left[ \frac{S}{cm \ln(K)} \right]$ | $B_{GM} \left[ \frac{S}{cm} \right]$ | $a$ [nm] | $g_T$ |
|---|---|---|---|---|---|---|
| 0 | 5 | 4 | 7.23 | -20.8 | 35.7 | 0.053 |
| 0 | 5 | 20 | 9.60 | 227 | 26.9 | 1.62 |
| 5 | 15 | 6 | 6.09 | 80.4 | 42.4 | 1.02 |
| 5 | 15 | 30 | 7.83 | 315 | 33.0 | 2.51 |
| 5 | 20 | 40 | 7.43 | 418 | 34.8 | 3.38 |

**Table S2. Fit parameters acquired from fitting temperature dependence of conductivity for samples fit to a granular metal conduction mechanism.**

$a$ is the metallic grain diameter and $g_T$ is the dimensionless tunneling conductance.

| Sn doping [%] | nominal diameter [nm] | In$_2$O$_3$ ALD cycles | $A_{FL}\left[\frac{S}{cm\,K^{1/2}}\right]$ | $B_{FL}\left[\frac{S}{cm}\right]$ |
|---|---|---|---|---|
| 0 | 5 | 4 | 1.45 | -1.54 |
| 0 | 5 | 20 | 2.41 | 248 |
| 5 | 15 | 6 | 2.65 | 89.1 |
| 5 | 15 | 30 | 2.69 | 328 |
| 5 | 20 | 0 | 1.29 | -1.60 |
| 5 | 20 | 49 | 2.38 | 431 |

**Table S3. Fit parameters acquired from fitting temperature dependence of conductivity for samples fit to a Fermi liquid conduction mechanism.**

| Sn doping [%] | nominal diameter [nm] | In$_2$O$_3$ ALD cycles | $\alpha\ [K^{-1}]$x10$^{-5}$ |
|---|---|---|---|
| 0 | 5 | 20 | 8.36 |
| 5 | 15 | 30 | 4.75 |
| 5 | 20 | 49 | 9.25 |

**Table S4. Fit parameters for conventional metals.**
$\alpha$ is the temperature coefficient of resistance (TCR).

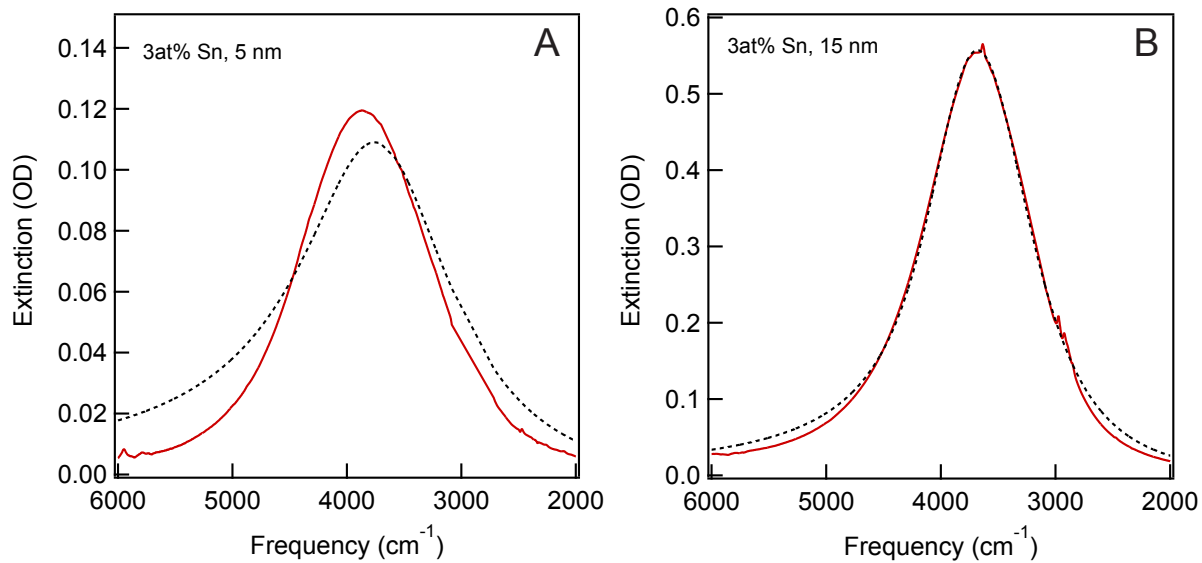

**Fig. S9. Fits to LSPR ITO nanocrystals in dispersion.**
Measured optical extinction (red solid line) fit to the HEDA model (black dashed line) for dilute nanocrystals dispersed in tetrachloroethylene. Fits for all other samples can be found in prior work.(*3*)

| Nominal Diameter (nm) | | 5 | 15 |
|---|---|---|---|
| **Nominal Doping (at%Sn)** | | **3** | **3** |
| Measured Values | $\mu_r$ (nm) | 3.8 | 6.7 |
| | $\sigma_r$ (nm) | 0.5 | 0.5 |
| | at% Sn | 3.22 | 3.83 |
| HEDA Fit Results | $\mu_{ne}$ (cm$^{-3}$) | 6.7E20 | 5.9E20 |
| | $\sigma_{ne}$ (cm$^{-3}$) | 3.2E19 | 7.6E19 |
| | $f_e$ | 0.30 | 0.46 |
| | $l_{bulk}$ (nm) | 17.0 | 17.0 |

**Table S5. Measured properties and HEDA fit results.**

$\mu_r$ and $\sigma_r$ are the mean nanocrystal radius and its standard deviation, $\mu_{ne}$ and $\sigma_{ne}$ are the mean electron concentration and its standard deviation, $f_e$ is the fraction of electron accessible volume, and $l_{bulk}$ is the bulk mean free path.

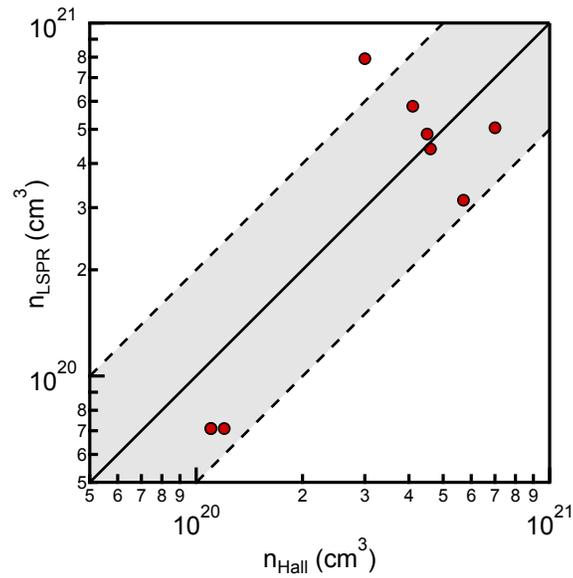

**Fig. S10. Comparison of *n* values derived from LSPR fits and Hall measurements.** Electron concentration values were extracted by LSPR fits ($n_{LSPR}$) and Hall effect measurement ($n_{Hall}$) for the exact same nanocrystal sample. Those values are compared in the plot above for 9 different samples. The solid line indicates where $n_{LSPR}$ is equal to $n_{Hall}$. The shaded region, within the dashed lines, indicates the area in which $n_{LSPR}$ and $n_{Hall}$ differ by less than a factor of two.

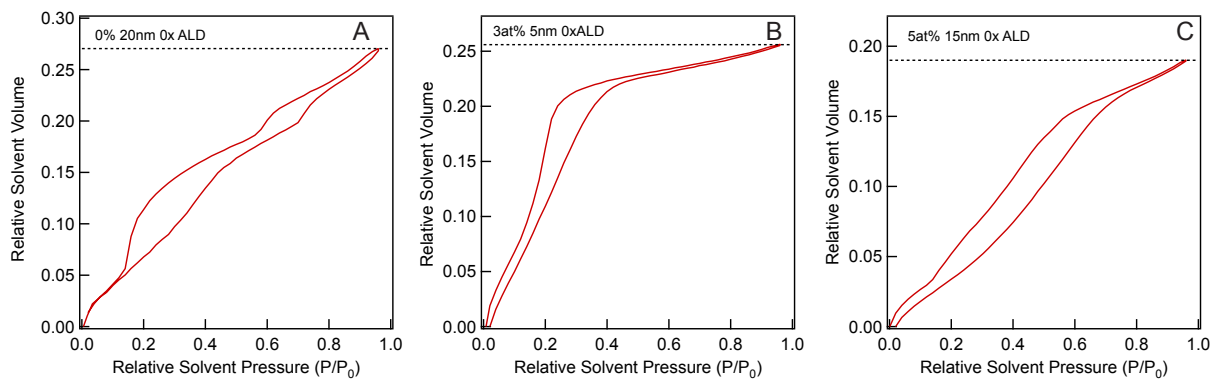

**Fig. S11. Ellipsometric Porosimetry (EP) of nanocrystal films.**
The maximum value of volume fraction (dashed line) denotes the fraction of void space (porosity) in our 0x indium oxide ALD samples, as shown for three nanocrystal films.